# Transition from chemisorption to physisorption of $H_2$ on Ti functionalized [2,2,2]paracyclophane: A computational search for hydrogen storage.

Rakesh K. Sahoo, Sridhar Sahu[*]

Computational Materials Research Lab, Department of Physics, Indian Institute of Technology (Indian School of Mines) Dhanbad, India

## Abstract

In this work, we studied the hydrogen adsorption-desorption properties and storage capacities of Ti functionalized [2,2,2]paracyclophane (PCP222) using density functional theory and molecular dynamic simulation. The Ti atom was bonded strongly with the benzene ring of PCP222 via Dewar interaction. Subsequently, the calculation of the diffusion energy barrier revealed a significantly high energy barrier of 5.97 eV preventing the Ti clustering over PCP222 surface. On adsorption of hydrogen, the first $H_2$ molecule was chemisorbed over PCP222 with a binding energy of 1.79 eV with the Ti metals. Further addition of $H_2$ molecules, however, exhibited their physisorption over PCP222-Ti through the Kubas-type $H_2$ interaction. Charge transfer mechanism during the hydrogen adsorption was explored by the Hirshfeld charge analysis and electrostatic potential map, and the PDOS, Bader's topological analysis revealed the nature of the interaction between Ti and $H_2$. The PCP222 functionalized with three Ti atoms showed a maximum hydrogen uptake capacity of up to 7.37 wt%, which was fairly above the US-DOE criterion. The practical $H_2$ storage estimation revealed that at ambient conditions, the gravimetric density of up to 6.06 wt% $H_2$ molecules could be usable, and up to 1.31 wt% of adsorbed $H_2$ molecules were retained with the host. The ADMP molecular dynamics simulations assured the reversibility by desorption of adsorbed $H_2$ and the structural integrity of the host material at sufficiently above the desorption temperature (300K and 500K). Therefore, the Ti-functionalized PCP222 can be considered as a thermodynamically viable and potentially reversible $H_2$ storage material.

## 1 Introduction

Extensive use of fossil fuels not only results in the depletion of those energy resources but also leads the world towards an alarming environmental catastrophe in terms of pollution and global warming. These consequences have motivated researchers across the globe to search for alternative sustainable and environment-friendly energy resources. Therefore, hydrogen drew the attention because it is considered as an ideal, pollution-free, and sustainable energy carrier, which can replace fossil fuels by fulfilling the energy need of the world, and thus can resolve the pollution due to fossil fuels[1, 2]. However, the major difficulty in hydrogen energy as fuel for domestic and vehicular application is its efficient storage and delivery at ambient conditions. Hydrogen can be stored mainly in two ways: system-based and material-based. System-based storage methods which is being adopted by few industries require huge volume vessels which should be made of composite material to withstand high pressure (~70 MPa) making the process quite expensive. However, compressed hydrogen storage systems are reported to have low volumetric densities, even at high pressure [3], and hydrogen storage in liquid state requires a very low temperature (~ -253°C) under high pressure (~ 250-350 atm) which is highly prone to safety concerns. On the other hand, the solid-state material-based hydrogen storage method is substantiated as efficient alternative to use hydrogen energy provided it adsorbs and desorb a desirable amount of $H_2$ at ambient conditions [4]. In solid-state materials, hydrogen is usually adsorbed by the physisorption or chemisorption process. In the physisorption process, the adsorbed hydrogen binds in molecular to the surface of host materials through weak interaction (adsorption energy ~ 0.1-0.8 eV/$H_2$). However, in the chemisorption process, the $H_2$ molecules dissociate into individual H atoms and migrate to the host materials by producing a strong chemical bond (with a binding energy of >1 eV/$H_2$) with the host atoms. Another type of adsorption process observed is similar to the physisorption, in which the inter-atomic H-H bond in the $H_2$ molecule is elongated but not dissociated and adsorbed by Kubas-type orbital interactions[2]. It enhances the $H_2$ adsorption energy and makes most of the $H_2$ storage capacities that fulfil the target of the US department of energy (DOE-US) [5, 6].

Since last few years, researchers are engaged extensively to study various materials, including carbon nanostructures [7, 8], metal hydrides [9, 10], graphene [11, 12], metal alloys[13, 14], metal-organic frameworks (MOF)[15, 16], and covalent-organic frameworks [17], etc. for the reversible hydrogen storage at ambient condition. However, it has been reported that these materials often have several limitations, including poor storage capacity, instability at significantly high temperatures, and low reversibility at normal temperatures. For example, Mg-based metal hydrides showed a high storage capacity of up to 7.6 wt% under ambient condition, however; it could be used only for 2-3 cycles [18].Similarly, metal alloys have very poor reversibility when used as hydrogen storage materials [19]. Using MOFs as $H_2$ storage materials, researchers could attain up to 15 wt% of storage capacity at temperatures and pressures of 77 K and 80 bar. However, under normal environmental condition its gravimetric and volumetric storage capacity remained very low [20]. To address the aforesaid issue and to develop commercially effective hydrogen storage materials, the experimentally synthesized organic compounds functionalized with transition metals (TMs), such as TM-doped organometallic buckyballs, TM-ethylene, etc., were introduced and investigated extensively [21, 22]. Early reports show that the TM atoms form a strong bond with the π-electron delocalized compounds through the Dewar mechanism and adsorb hydrogen molecules via Kubas interaction[23, 24]. For example, Chakraborty *et al*. studied the hydrogen storage in Ti-doped Ψ-graphene and reported an $H_2$ uptake capacity of up to 13.1 wt% with an average adsorption energy of -0.30 eV/$H_2$ [25]. Dewangan *et al*. predicted up to 10.52 wt% of $H_2$ adsorption in Ti-functionalized holey graphyne via the Kubas mechanism with adsorption energy and desorption temperature of 0.38 eV/$H_2$ and 486 K, respectively[26].

Numerous theoretical and experimental studies revealed that metal-adorned small organic molecules like $C_nH_n$ could capture a large number of $H_2$ molecules. For example, Zhou *et al*. estimated hydrogen uptake capacity up to 12 wt% in $TiC_2H_4$ with $H_2$ binding energy of 0.24 eV/$H_2$[27, 28]. High capacities of $H_2$ storage in $TMC_2H_4$ (M = Ti, Sc, V, Ni, Ce, Nb) complexes was reported by Chaudhari *et al*.[29, 30, 31]. At low benzene pressure (35 millitorrs) and ambient temperature, $TiC_6H_6$ was experimentally shown to absorb up to 6 wt% hydrogens [32]. Phillips *et al*. obtained an $H_2$ uptake of up to 14 wt% and quick kinetics at room temperature on $TiC_6H_6$ by laser ablation; however, the experiments did not discuss the desorption process[33]. Recently, Ma *et al.* theoretically studied an interesting combination of chemisorption and physisorption in Ti-doped $C_6H_6$ and reported an uptake capacity of 6.02 wt % with complete desorption at 935 K [34]. Mahamiya *et al.* revealed the $H_2$ storage capacities of 11.9 wt % in K and Ca decorated biphenylene with an average adsorption energy of 0.24-0.33 eV [35]. Y atom doped zeolite showed high capacity adsorption of $H_2$ with binding energy 0.35 eV/$H_2$ and the desorption energy of 437K for fuel cells[36].

Macrocyclic compounds, like paracyclophane (PCP), a subgroup derivative of cyclophanes, comprises aromatic benzene rings with number of -$CH_2$- moieties linking the subsequent benzene rings [37]. The PCPs are easier to synthesize in the laboratory; they can be functionalized with metal atoms due to the presence of aromatic benzene rings in the geometry, making them a feasible alternative for hydrogen storage prospects. For instance, Sathe *et al.* studied the Sc and Li decorated PCP and reported the molecular $H_2$ physisorbed via Kubas-Niu-Jena interaction resulting in up to 10.3 wt% $H_2$ uptake capacity [38]. The hydrogen storage transition metal (Sc, Y) functionalized [1,1]paracyclophane was investigated by Sahoo *et al.* and reported a storage capacity of 6.33-8.22 wt%, with an average adsorption energy of 0.36 eV/$H_2$ and desorption temperature of 412 K - 439 K[39]. The $H_2$ storage on Li and Sc functionalized [4,4]paracyclophane shows an uptake capacity of 11.8 wt% and 13.7 wt%, as estimated by Sathe *et al.* [40]. Kumar *et al.* revealed the combination of physisorption and chemisorption of hydrogen on Sc and Ti functionalized BN-analogous [2.2]PCP[41]. They showed the first hydrogen molecule chemisorbed on the host material followed by physisorption of other $H_2$, resulting in a storage of ~8.9 wt% via Kubas interaction. Numerous other metal-decorated macrocyclic compounds have been explored as hydrogen storage possibilities, with storage capacities above the DOE requirement; however, only a few have shown practical $H_2$ capacity at varied thermodynamic conditions. Though few PCP-based hydrogen storage systems are available in the literature, the [2,2,2]paracyclophane, which is experimentally synthesized by Tabushi *et al.*[42] is yet to be explored as a hydrogen storage material.

In the present work, we investigated the chemisorption and physisorption properties of hydrogen molecules on [2,2,2]paracyclophane (PCP222) functionalized with Ti atoms and estimated their hydrogen uptake capacity at varied thermodynamics. In paracyclophane, there are many molecules in the group and are named after their pattern of arene substitution. The preceding square bracket number, "[2,2,2]" in [2,2,2]paracyclophane, indicates that the consecutive benzene rings (3 benzene rings) in paracyclophane are linked with two (-$CH_2$-) moieties. The linking bridges are relatively short; thus, the separation between consecutive benzene rings is small, which develops a strain in the aromatic rings. This strain in the rings can be utilized for Ti functionalization over the aromatic benzene ring. Due to the strain and metal functionalization, the aromatic benzene rings lose their inherent planarity. We choose to functionalize Ti metal atoms over the PCP222, as the d- block transition metal elements are well known for reversible hydrogen adsorption and could bind the $H_2$ molecules via Kubas interaction[25, 26]. Though there are few reports available based on hydrogen storage in macrocyclic organic compounds and other Ti-doped nanostructures, our work is the first to investigate the efficiency of Ti-functionalized PCP222 using the atomistic MD simulation, practical storage capacity, and diffusion energy barrier estimation

## 2 Theory and Computation

We have performed the theoretical calculations on [2.2.2] paracyclophane (PCP222) and their hydrogenated structures within the framework of density functional theory (DFT)[43]. In the computation, the advanced hybrid ωB97Xd functional is used, and molecular orbitals (MO) are expressed as the linear combination of atom-centered basis function for which the valence diffuse and polarization function 6-311+G(d,p) basis set is used for all atoms. ωB97Xd includes the long-range and Grimme's D2 dispersion correction which is a range-separated version of Becke's 97 functional[44, 45]. It is important to note that the ωB97Xd technique is a trustworthy method for studying the non-covalent interactions, Organometallic complexes, and their thermochemistry. To ensure the studied structures are in true ground state on the potential surface, the harmonic frequencies of all the systems are determined and are found to be positive. All the theoretical computations are performed with the computational program Gaussian 09[43].

In order to investigate the binding strength of titanium (Ti) atoms on the PCP222, we have calculated the average binding energy of decorated Ti atoms by using the following equation.

$$E_b = \frac{1}{m}[E_{PCP222} + mE_{Ti} - E_{PCP222+mTi}] \quad (1)$$

Where $E_{PCP222}$, $E_{Ti}$, and $E_{PCP222+mTi}$ is the total energy of PCP222, Ti atom and Ti-decorated PCP222 respectively. m is the number of Ti atoms added PCP222.

The average adsorption energy of molecular hydrogen with metal atoms is calculated as[46].

$$E_{ads} = \frac{1}{n}[E_{PCP222+mTi} + nE_{H_2} - E_{PCP222+mTi+nH_2}] \quad (2)$$

Where $E_{PCP222+mTi}$, $E_{H_2}$, and $E_{PCP222+mTi+nH_2}$ is the total energy of host material, hydrogen molecule and hydrogen trapped complexes respectively. n is the number of $H_2$ molecules adsorbed in each complex.

The global reactivity descriptors such as hardness (η), electronegativity (χ), and electrophilicity (ω) were estimated and used to study the stability and reactivity of Ti functionalized PCP222 and their hydrogen adsorbed derivatives [47, 48]. The energy gap between the highest occupied molecular orbital (HOMO) and lowest unoccupied molecular orbital (LUMO) is computed to assure the kinetic stability of the studied systems. Further, to understand the electronic charge transfer properties, the Hirshfeld charge and electrostatic potential map (ESP) were explored. Moreover, partial density of states (PDOS) investigation was also carried out to further understand the process of hydrogen interaction. The topological parameters were studied using Bader's theory of atoms in molecules (AIM) to analyze more about the nature of the interaction between metal on PCP222 and adsorbed hydrogen molecules.

To obtained the hydrogen uptake capacity, gravimetric density (wt%) of hydrogen is calculated using the following equation[49]:

$$H_2(wt\%) = \frac{M_{H_2}}{M_{H_2} + M_{Host}} \times 100 \quad (3)$$

Here $M_{H_2}$ represent the mass of the total number of $H_2$ molecules adsorbed and $M_{Host}$ represent the mass of metal-doped PCP222.

## 3 Results and Discussion

### 3.1 Structural properties of PCP222

The optimized geometrical structure of PCP222 is depicted in Figure 1(a). PCP222 has three benzene rings connected by two -$CH_2$- moiety as a bridge between the consecutive rings. The distance between the two consecutive -$CH_2$- moiety and the -$CH_2$- across the benzene ring are found to be 1.54 Å and 5.84 Å respectively, which is consistent with the earlier experimentally reported value by Cohen-Addad *et al*. [50]. To validate the **π** aromaticity of the optimized molecule, we computed the Nucleus Independent Chemical Shift (NICS) of PCP222 before functionalizing by any metal atom. The NICS values are determined with 1 Å increment from the center to 3 Å above the three benzene rings. NICS(1) is found to be negative maximum (-10.1 ppm), suggesting the aromatic nature of PCP222. This indicates that the benzene rings of PCP222 are **π** electron-rich and can bind a metal atom outside the benzene rings.

### 3.2 Functionalization of Ti atom on PCP222

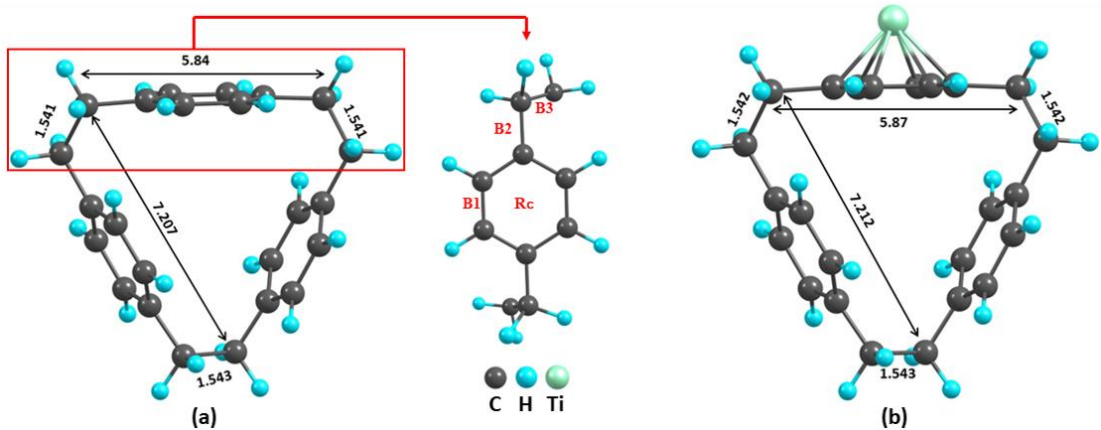

**Figure 1:** (a) Optimized structure of PCP222 with all possible marked adsorption site marked, (b) Ti functionalized PCP222

Next, we explore different possible adsorption sites of pristine PCP222, such as C-C bridge of benzene ring (B1), $CH_2$ moiety and benzene bridge (B2), $CH_2$ - $CH_2$ bridge (B3), and above the center of benzene ($R_c$) which are depicted in Figure 1(a). To design the host material for hydrogen adsorption, a single Ti atom is positioned about 2 Å above at the regioselective sites of PCP222, and the resulting structure is re-optimized. The binding energy between Ti and PCP222 calculated using Equation 1 at different adsorption sites shows that the Ti atom is stable at two positions, B3 and $R_c$ sites of PCP222 with binding energies of 0.37 eV and 2.20 eV, respectively which fairly agree with the previously reported value of Ti on CNT by Yildirim *et al.* [51]. Hence, the most favourable site for Ti atom functionalization is at the $R_c$ site above the benzene ring of PCP222.

### 3.2.1 Bonding mechanism of Ti on PCP222

To understand the binding mechanism of Ti on PCP222, we analyzed the partial density of state (PDOS), electrostatic potential map (ESP), Hirshfeld charge, and Bader's topological parameters of the Ti functionalized PCP222 system as discussed below.

**Density of states**

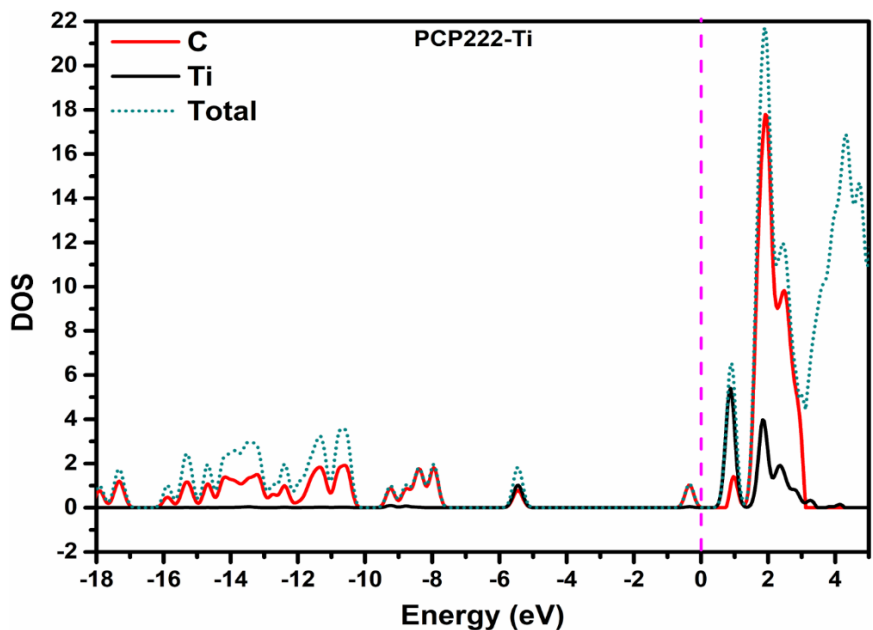

**Figure 2:** Density of states plot on Ti and C atom on PCP222

The Ti atom is functionalized on PCP222 via the Dewar mechanism in which $\pi$-electron gets transferred from the highest occupied molecular orbitals (HOMO) of the substrates to the vacant d-orbital of Ti followed by the back-donation of charges from the partially filled d-orbital of Ti to empty $\pi^*$-anti-bonding of the benzene ring of PCP222[26]. To understand the orbital interaction between the Ti and C atom of PCP222, we have performed the partial density of states (PDOS) calculation of PCP222-Ti and the result is plotted in Figure 2. Figure 2 clearly shows that the electronic states of the Ti atom and the C atom of PCP222 overlap below and above the Fermi level (E = 0). The transferred electrons partially fill the unoccupied states of PCP222, as seen by the intense peaks near the Fermi level. This infers an orbital interaction between Ti and C atom of PCP222 mediated by charge transfer. The fact is also obvious because Ti has the relatively lower ionization potential than the C atom.

**ESP and Hirshfeld charges**

To get a picture of electronic charge distribution over the PCP222 during Ti functionalization, we plotted the electrostatic potential (ESP) map over the total electron density, as shown in Figure.S1. The variation of electron density in the ESP map is shown by using different colour codes, which follows the pattern of accumulation and reduction of electron density as; red (maximum electron density) >orange > yellow > green > blue (minimum electron density). In the ESP plot (Figure.S1), the red region over the benzene ring of PCP222 implies the aggregation of electron density. After the functionalization of the Ti atom, the region changed to dark blue, indicating the deficiency of electron density around the metal making it susceptible to bind with the guest molecules. Moreover, region around the carbon atoms of PCP222 turns from red to green supporting the charge transfer as discussed above. The estimated Hirshfeld charge on C and Ti atoms is computed to be -0.121 e.u and +0.511 e.u, respectively, which makes the Ti atom nearly ionic, opening the possibility for $H_2$ adsorption.

### 3.2.2 Diffusion energy barrier calculation

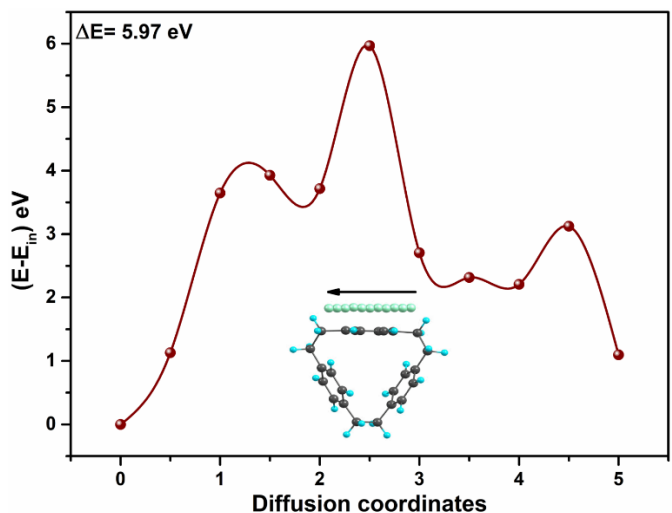

**Figure 3:** Ti diffusion energy barrier over the PCP222

According to earlier reports, the aggregation of transition metal atoms on the substrate may lower the ability of the host material for hydrogen adsorption. So, before hydrogen adsorption on the surface of PCP222, it is necessary to study the possibility of metal clustering on the substrate. If the Ti atom is displaced from its stable adsorption position on PCP222 due to an increase in temperature, there is a strong possibility of metal clustering. Since the Ti binding energy on PCP222 (2.2 eV/Ti) is lower than the cohesive energy of an isolated single Ti atom (4.85 eV), we evaluated whether or not there is an energy barrier for Ti atom diffusion on PCP222. The diffusion energy barrier is calculated by displacing Ti to a finite neighbourhood ($\delta r$) over the adsorption site of PCP222. as shown in Figure 3. The difference in energy calculated between the initial and that of the close neighbourhood is then plotted with the diffusion coordinates as shown in Figure 3. The figure illustrates the diffusion energy barrier to be 5.97 eV, which is sufficient to prevent the diffusion of the Ti atom over PCP222 and therefore avoid Ti-Ti clustering which is also supported by the works of Dewangan *et al.* [26] and Chakraborty *et al.* [25]. Therefore, the present Ti functionalized PCP222 can be considered a suitable candidate for hydrogen adsorption.

### 3.3 Adsorption of H$_2$ molecules on PCP222-Ti

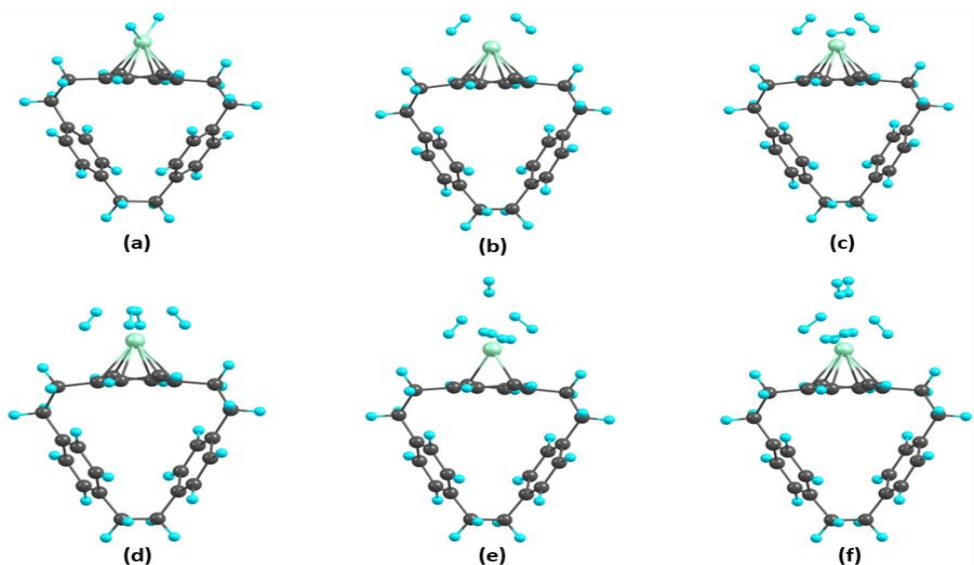

**Figure 4:** Optimized geometry of hydrogenated Ti functionalized PCP222, (a) PCP222-Ti-1H$_2$, (b) PCP222-Ti-2H$_2$, (c) PCP222-Ti-3H$_2$, (d) PCP222-Ti-4H$_2$, (e) PCP222-Ti-5H$_2$, (f) PCP222-Ti-6H$_2$

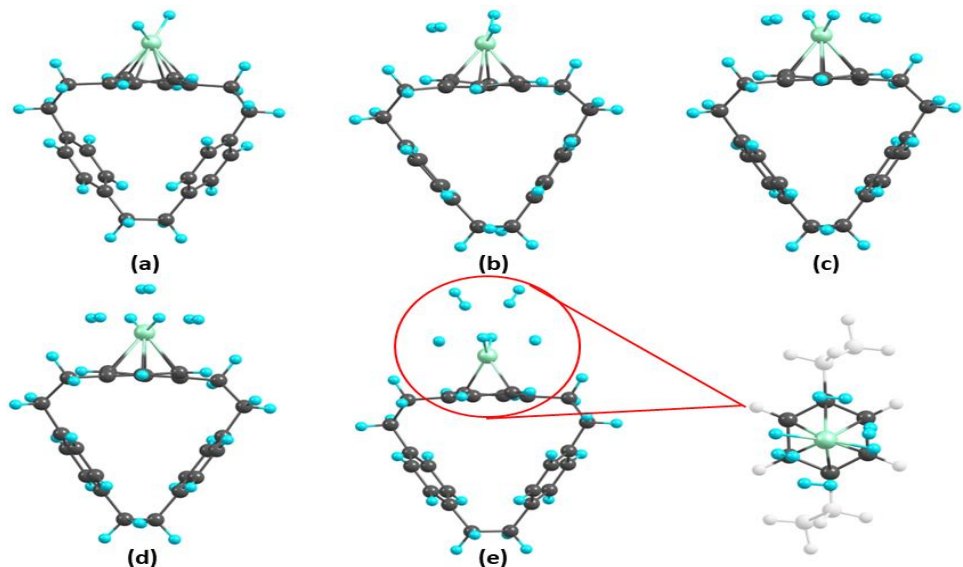

**Figure 5:** Optimized geometry of hydrogenated Ti functionalized PCP222, (a) PCP222-Ti-2H, (b) PCP222-Ti-2H-1H$_2$, (c) PCP222-Ti-2H-2H$_2$, (d) PCP222-Ti-2H-3H$_2$, (e) PCP222-Ti-2H-4H$_2$.

**Table 1:** Average bond distance between carbon bridge (C-C), center of PCP222 benzene ring ($R_c$) and Titanium atom ($R_c$-Ti), Titanium and hydrogen molecules (Ti-$H_2$), and hydrogen Hydrogen (H-H) in Å. Average adsorption energy of $H_2$ on PCP222-Ti.

| Name of complex | Bridge C-C | $R_c$ -Ti | Ti-H | H-H | $E_{ads}$ (eV) |
|---|---|---|---|---|---|
| PCP222-Ti | 1.542 | 1.566 | | | |
| PCP222-Ti-2H | 1.540 | 1.800 | 1.750 | 2.796 | 1.797 |
| PCP222-Ti-2$H_2$ | 1.540 | 1.765 | 1.770 | 0.884 | 0.953 |
| PCP222-Ti-3$H_2$ | 1.540 | 1.798 | 1.830 | 0.852 | 0.784 |
| PCP222-Ti-4$H_2$ | 1.540 | 1.818 | 1.905 | 0.806 | 0.672 |
| PCP222-Ti-5$H_2$ | 1.540 | 1.842 | 2.332 | 0.816 | 0.554 |
| PCP222-Ti-6$H_2$ | 1.540 | 1.842 | 2.633 | 0.804 | 0.467 |
| PCP222-Ti-2H-1$H_2$ | 1.540 | 1.822 | 1.926 | 0.800 | 0.480 |
| PCP222-Ti-2H-2$H_2$ | 1.540 | 1.837 | 1.868 | 0.803 | 0.474 |
| PCP222-Ti-2H-3$H_2$ | 1.540 | 1.851 | 1.899 | 0.801 | 0.406 |
| PCP222-Ti-2H-4$H_2$ | 1.540 | 1.837 | 2.840 | 0.774 | 0.256 |

To investigate the hydrogen adsorption on the surface of Ti functionalized PCP222, we added the $H_2$ molecules sequentially to PCP222-Ti. First, we added a single $H_2$ molecule at about 2 Å above the Ti atom functionalized on PCP222 and allowed the system to relax. It is observed that the $H_2$ molecule dissociates into two fragments of H atoms and forms chemical bond with the Ti atom. The Ti-H bond length is found to be 1.75 Å which is close to the experimental result for titanium monohydride [52]. The H-H bond distance is noted to be about 2.8 Å (Figure 4(a)). The binding energy between Ti and H is calculated to be 1.79 eV which lies in the range of chemisorption mechanized by Kubas's interaction [2, 38]. Similar result was also reported by Ciraci *et al*. for the adsorption of a single $H_2$ molecule on Ti-decorated SWNT8 ( and SWBNT ) where the $H_2$ molecules dissociate into individual H atoms with a binding energy of 0.83 eV/H (0.93 eV/H) and H-H- distance of 2.71Å (3.38 Å)[51, 53]. However, when two $H_2$ molecules are simultaneously added to the sorption center, the calculated average adsorption energy is reduced to 0.95 eV/$H_2$, with the average H-H bond length stretching from 0.74 Å to 0.8 Å. This result clearly indicates the adsorption process to be physisorptive. This is because of reduced interaction strength between Ti atoms and $H_2$ molecules caused due to screening effect. From the ESP analysis (7) it is obvious that simultaneous presence of two $H_2$ molecules reduces the charge densities of Ti and $H_2$ thereby inducing a weak charge polarization which causes the physisorption of hydrogen on the surface of Ti functionalized PCP222. Another way of generating similar isomeric configuration is chemisorption induced physisorption of $H_2$ molecules on Ti functionalized PCP222 in which one $H_2$ molecule is adsorbed over n PCP222-Ti-2H (Figure 5(b)). Interestingly, this configuration is 0.37 eV lower in energy than that of PCP222-Ti-$2H_2$, and the $H_2$ adsorbed with lower adsorption energy (0.48 eV). Therefore, we proceed with both configurations for further hydrogen adsorption. Sequential adsorption of $H_2$ molecules on PCP222-Ti results in the maximum adsorption up to $6H_2$ molecules. The adsorption of 3rd, 4th, 5th, and 6th $H_2$ molecules to PCP222-Ti reduces the average $H_2$ adsorption energy to 0.784, 0.68, 0.554, and 0.467 eV/$H_2$, respectively. On the other hand, successive addition of $H_2$ molecules to PCP222-Ti-2H leads to maximum adsorption of four hydrogen molecules. More addition of $H_2$ molecules beyond maxima in both the cases causes them to fly away from the sorption center. It is observed that the average adsorption energy decreases with an increase in the number of $H_2$ molecules in the system which is due to the steric hindrance among the adsorbed $H_2$ crowed and the increase in distances between the $H_2$ and sorption centers. The estimated data of adsorption energy and geometrical parameters of all the bare hydrogenated systems and presented in Table 1.

### 3.3.1 Partial density of states

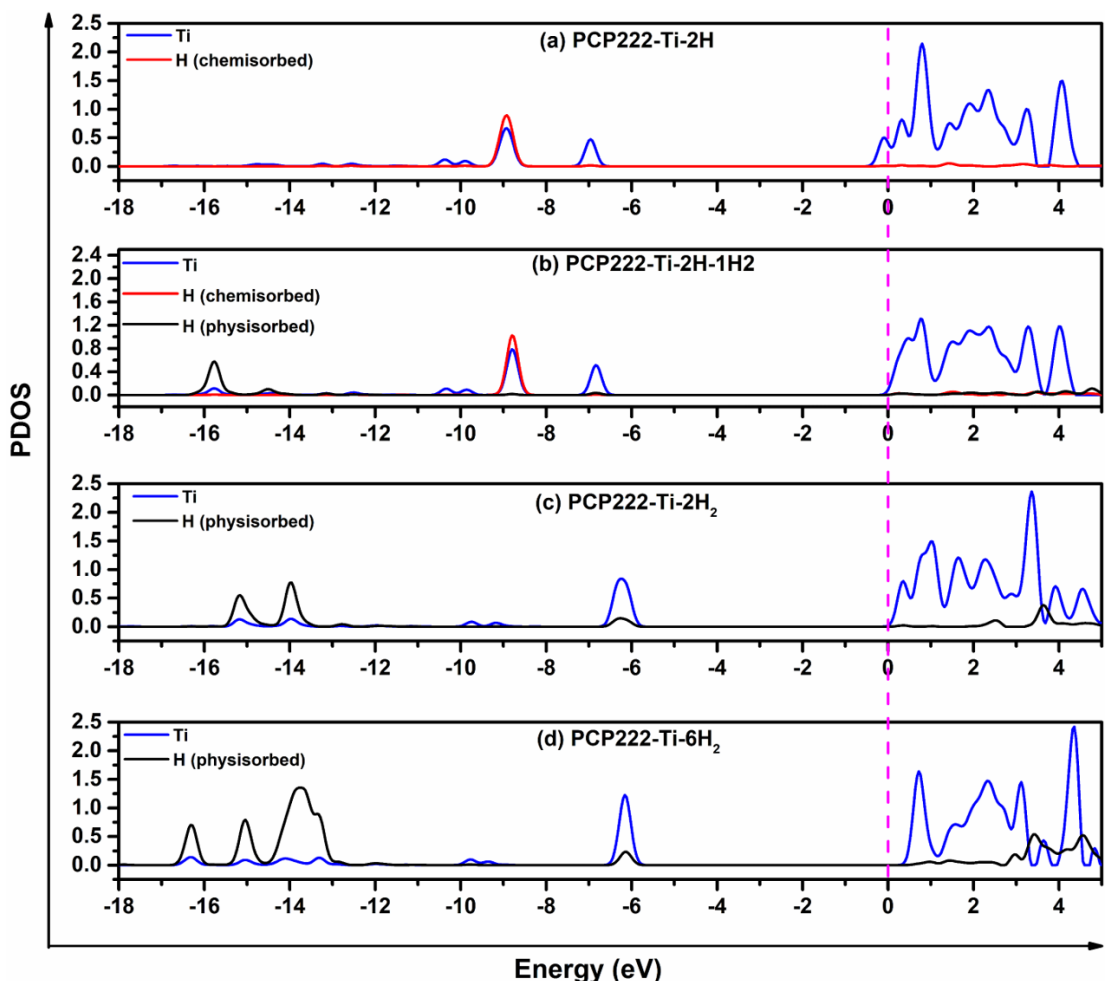

**Figure 6:** Partial density of state on Ti and H atoms of (a) PCP222-Ti-2H, (b) PCP222-Ti-2H-1H$_2$, (c) PCP222-Ti-2H$_2$, and (d) PCP222-Ti-6H$_2$

The partial density of states (PDOS) of Ti and H atoms of the hydrogen adsorbed PCP222-Ti with the chemisorbed, and physisorbed hydrogen is plotted in Figure 6. The adsorption of 1H$_2$ to the host resulting in chemisorption is contributed from the strong overlapping of H and Ti orbital near -9 eV. Upon adsorption of another H$_2$ molecule over PCP222-Ti-2H, the peaks of $\sigma$-orbital (HOMO) of hydrogen and Ti orbital appears at around -15.7 eV below the Fermi level and $\sigma^*$ (LUMO) of hydrogen interacts with the orbital of Ti and chemisorbed H above the Fermi level (figure 6(b)) which can be explained by the Kubas mechanism in which a small charge transfer occurs from the $\sigma$(HOMO) orbital of H$_2$ to the vacant 3d orbital of the Ti atom, followed by a back-donation of charges in the other direction from the partially filled 3d orbitals of Ti to $\sigma^*$ (LUMO) of H$_2$ molecules. When two H$_2$ molecules are introduced simultaneously to the PCP222-Ti, similar DOS peaks are observed, suggesting the H$_2$ adsorption via the Kubas mechanism. However, here the σ orbital of H$_2$ splits into several peaks in the range of -15.2 to -6.2 eV and moves closer to the Fermi level inferring lower in the interaction strength. On adsorption of 6H$_2$ molecules to Ti functionalized PCP222, the σ orbitals split into numerous peaks in a broad range of -16.3 eV to -6.1 eV with enhanced intensity. This signifies that the adsorption strength gets weaker with an increase in the quantity of H$_2$ molecules in the host systems.

### 3.3.2 Electrostatics potential and Hirshfeld charges

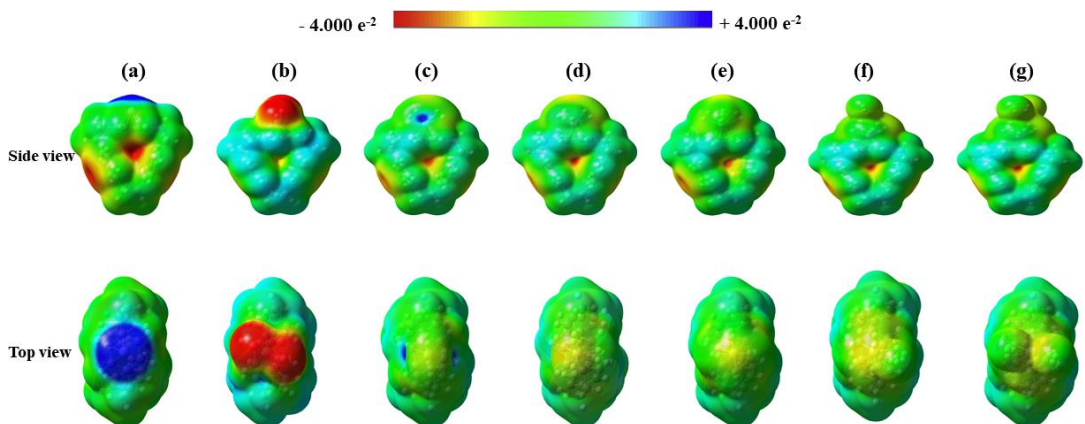

**Figure 7:** Electrostatics potential map of (a) PCP222-Ti, (b) PCP222-Ti-2H, (c) PCP222-Ti-2H$_2$, (d) PCP222-Ti-3H$_2$, (e) PCP222-Ti-4H$_2$, (f) PCP222-Ti-5H$_2$, (f) PCP222-Ti-6H$_2$.

To obtain a qualitative depiction of electronic charge distribution over the bare and hydrogenated PCP222-Ti, we generated and plotted the electrostatic potential (ESP) map on the total electron density as shown in Figure 7 The charge distribution is used to determine the active adsorption region for the guest hydrogen molecules. The dark blue zone above the Ti atom on PCP222-Ti (Figure 7(a)) and the dark red region over the first adsorbed hydrogen atom indicates a strong interaction between them leading to chemisorption of hydrogen atom. Upon adsorption of two H$_2$ molecules simultaneously, the region over Ti turns from dark blue to light blue, suggesting the fact that, positive charge get transferred from the Ti atom to the adsorbed H$_2$ and C atom of PCCP222 thereby inducing charge polarization which causes physisorption of the second H$_2$ molecule. Further addition of H$_2$ molecules to PCP222-Ti, the region over Ti atom turns to bluish-green and then to green inferring further charge transfer (depletion of electron density near Ti ) and the yellow region over the adsorbed H$_2$ represents a little accumulation of electron density at hydrogen molecules[26].

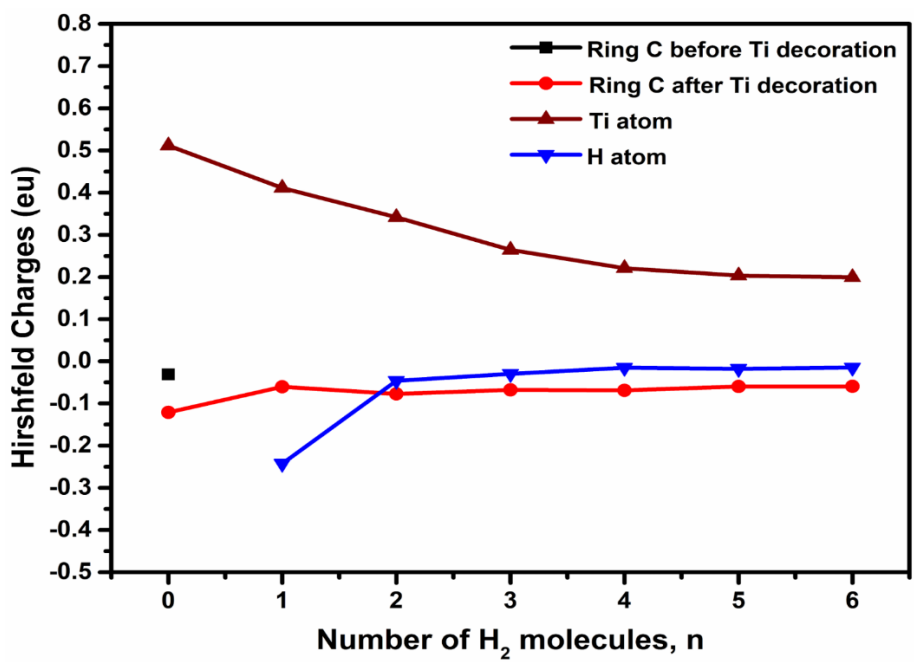

**Figure 8:** Hirshfeld charges before and after hydrogen adsorption on PCP222-Ti

Figure 8 shows the average Hirshfeld charges on the Ti atom, the adsorbed $H_2$ molecules, and the C atoms of the benzene ring (Ti functionalized site) as a function of the number of $H_2$ adsorbed on the host. The average charges on the C atom of the benzene ring are initially computed to be -0.031 e which then raises to -0.121 e with the functionalization of the Ti atom. The charge on the Ti atom of PCP222-Ti is found to be +0.511 e, indicating the transfer of electronic charges from the Ti atom to the C atom of the benzene ring. On chemisorption of the first hydrogen on PCP222-Ti, the electronic charges on the Ti and H atoms are +0.41 a.u and -0.24 a.u implying a strong attractive interaction between them as discussed above. Adding more $H_2$ molecules gradually lessen the Hirshfeld charges over the Ti and H atoms implying polarization induced weak interaction between them. (Figure 8).

### 3.3.3 Bader's topological analysis

The topological analysis at the bond critical point (BCP) is used to investigate the nature of interactions between the Ti-functionalized PCP222 and the adsorbed $H_2$ molecules employing Bader's quantum theory of atoms in molecules (QTAIM). The topological descriptors associated with the electronic distribution, such as electron density ($\rho$), Laplacian ($\nabla^2\rho$), and total energy density ($\mathcal{H}$) (calculated as the sum of local kinetic $G(\rho)$ potential energy density $V(\rho)$ ), at BCPs are presented in Table S1. Kumar *et al.* reported that the positive value of the Laplacian of electron density ($\nabla^2\rho > 0$) at BCP indicates a decrease in $\rho$ at the bonding region, suggesting an interaction of closed-shell (non-covalent) type [56]. For PCP222-Ti-6H$_2$, the value of $\rho$ and $\nabla^2\rho$ at BCP of Ti and adsorbed H$_2$ are found to be 0.057 a.u and 0.208 a.u, respectively which infers a closed-shell interaction between Ti and H$_2$. Moreover, the negative value of $\mathcal{H}_{BCP}$ and $-\frac{G(\rho)}{V(\rho)} > 1$ at BCP of Ti and H$_2$ confirm the closed-shell interaction among sorption center and H$_2$ as proposed by Koch *et al.* (Table S1) [57]. For C–C and C-Ti bond, the average $\rho$ value shows very nominal changes after the hydrogen adsorption which suggests the post-adsorption chemical stability of the host material. Additionally, the average $\rho$ on BCP of the H-H bond in PCP222-Ti-6H$_2$ is 0.231 a.u which is almost the same as on isolated bare H$_2$ molecule (0.263 a.u). This implies that the adsorbed hydrogens are in quasi-molecular form during the adsorption which also reflected in H-H bond elongation by 0.06-0.14 Å.

### 3.4 Thermodynamically usable H$_2$ capacity

### 3.4.1 Storage capacity

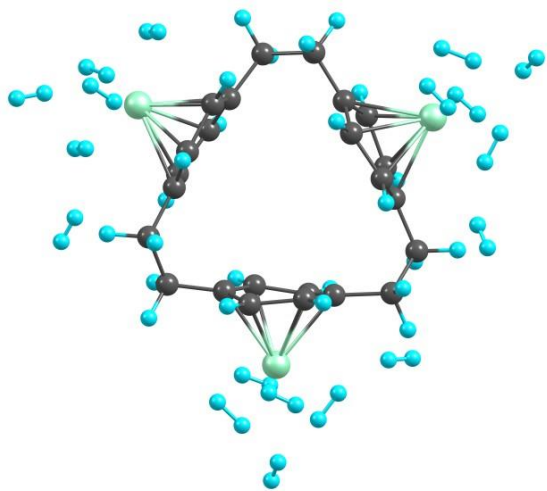

**Figure 9:** Optimized geometry of hydrogen saturated 3Ti functionalized PCP222

To examine the maximum H$_2$ gravimetric storage capacity of the system, we have functionalized the Ti atom on each benzene ring of PCP222 resulting in the structure of PCP222-3Ti as shown in Figure 9 and S3. Further, we added H$_2$ molecules to each Ti atom functionalized on PCP222 sequentially as discussed in previous section (3.3). The calculated average $H_2$ adsorption energy and the change in geometrical parameters are presented in Table 2. The adsorption of $H_2$ on PCP222-3Ti is observed to behave similar to that of on single Ti atom on PCP222. On saturation of the $H_2$ uptake capacity of PCP222-3Ti, each sorption center is found holding a maximum of $6H_2$ molecules with a gravimetric storage capacity of 7.37 wt%. Since the first $H_2$ molecule on each Ti atom dissociate into two H atom and bonded strongly with Ti atoms, 1.31 wt% of hydrogen adsorbed via the chemisorption process is difficult to desorb. However, the concurrent addition of two or more $H_2$ molecules to each Ti atom over PCP222, results in physisorption kind of adsorption. Further, to confirm the stability of maximum hydrogenated systems, the energy gap ($E_g$) (gap between HOMO-LUMO) and global reactivity parameters such as $\eta, \chi,$ and $\omega$ were estimated using the Koopmans theorem[58]. Notwithstanding, the studied system follow the "*maximum hardness and minimum electrophilicity principle*," ensuring their chemical stability (Figure S4)[59].

**Table 2:** Average bond distance between carbon bridge (C-C), center of PCP222 benzene ring ($R_c$) and Titanium atom ($R_c$-Ti), Titanium and hydrogen molecules (Ti-$H_2$), and hydrogen-hydrogen (H-H) in Å. Average adsorption energy and successive desorption energy of PCP222-3Ti-$nH_2$ (n=3,6,9,12,15,18)

| Name of complex | Bridge C-C | $R_c$-Ti | Ti-H | H-H | $E_{ads}$ (eV) | $E_{des}$ (eV) |
|---|---|---|---|---|---|---|
| PCP222_3Ti | 1.543 | 1.590 | - | - | - | - |
| PCP222_3Ti-3$H_2$ | 1.537 | 1.799 | 1.747 | 2.824 | 1.824 | 1.824 |
| PCP222_3Ti-6$H_2$ | 1.537 | 1.756 | 1.776 | 0.880 | 0.988 | 0.152 |
| PCP222_3Ti-9$H_2$ | 1.537 | 1.790 | 1.832 | 0.849 | 0.813 | 0.464 |
| PCP222_3Ti-12$H_2$ | 1.536 | 1.824 | 1.801 | 0.821 | 0.700 | 0.360 |
| PCP222_3Ti-15$H_2$ | 1.535 | 1.825 | 2.332 | 0.806 | 0.570 | 0.050 |
| PCP222_3Ti-18$H_2$ | 1.536 | 1.838 | 2.622 | 0.803 | 0.482 | 0.043 |

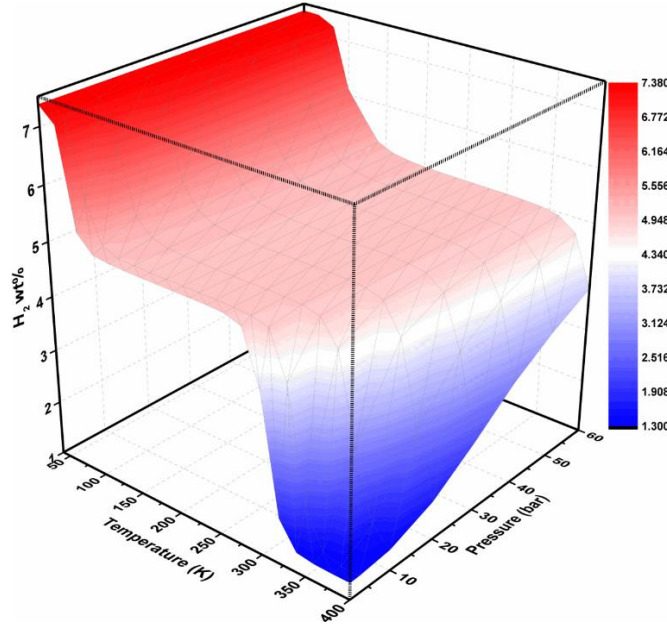

**Figure 10:** Hydrogen occupation number for PCP222-3Ti at various T and P.

For a practically usable hydrogen medium, a substantial amount of $H_2$ molecules should be adsorbed by the host material at attainable adsorption conditions and the adsorbed $H_2$ molecules should be desorbed effectively at a suitable temperature (T) and pressure (P). Thus, it is essential to estimate the number of hydrogen molecules usable at a wide variety of T and P. We have estimated the usable hydrogen gravimetric density of the studied system by calculating the number of $H_2$ molecules stored in PCP222-3Ti at different T and P using the empirical value of $H_2$ gas chemical potential (μ). The $H_2$ gravimetric density is estimated from the occupation number (N) by the following equation and plotted with various T and P in Figure 10[60].

$$N = \frac{\sum_{n=0}^{N_{max}} n g_n e^{[n(\mu - E_{ads})/K_B T]}}{\sum_{n=0}^{n_{max}} g_n e^{[n(\mu - E_{ads})/K_B T]}} \quad (4)$$

Here $N_{max}$ is the maximum number of $H_2$ molecules adsorbed on each Ti atom on PCP222, n and $g_n$ represents the number of $H_2$ molecules adsorbed and configurational degeneracy for a *n* respectively. $k_B$ is the Boltzmann constant and -$E_{ads}$ (>0) indicates the average adsorption energy of $H_2$ molecules over PCP222-3Ti. $\mu$ is the empirical value of chemical potential of $H_2$ gas at specific T and P, obtained by using the following expression [61].

$$\mu = H^0(T) - H^0(0) - TS^0(T) + K_B T \ln\left(\frac{P}{P_0}\right) \quad (5)$$

Here $H^0(T)$, $S^0(T)$ are the enthalpy and entropy of $H_2$ at pressure $P_0$ (1 bar).

From the Figure 10 it is clear that, the PCP222-3Ti can store 18$H_2$ molecules at temperatures up to 80 K and 10-60 bar pressure. Up-to these thermodynamic conditions, the maximum $H_2$ storage capacity of the studied system is estimated as 7.37 wt%, which is consistent the experimentally reported value for Pd functionalized carbon nanotubes [62] and is fairly above the target set by US-DOE (5.5 wt% by 2025). On raising the temperature above 80 K, the $H_2$ molecules start to desorb from the PCP222-3Ti and retain >5.5 wt% of $H_2$ till the temperature of 120 K under 30-60 bar. Further, rise in temperature, the system maintains an $H_2$ gravimetric density of 5 wt% (close to the target of US-DOE) throughout a temperature range of 120-300 K and a pressure range of 3-60. This thermodynamic condition may be treated as an ideal storage condition for $H_2$ on PCP222-3Ti. At the temperature of 400 K and pressure of 1-10 bar, the system retains 1.31 wt% of hydrogen, that are adsorbed via the chemisorption process and may be desorbed at very high temperatures. Thus, a total gravimetric density of 6.06 wt% (difference in G.D at 80 K and 400 K) $H_2$ molecules are usable under ambient conditions, which is fairly higher than the US-DOE target. This result justifies that the Ti functionalization over PCP222 can be used as a potential reversible hydrogen storage material.

### 3.5 Molecular dynamics simulations

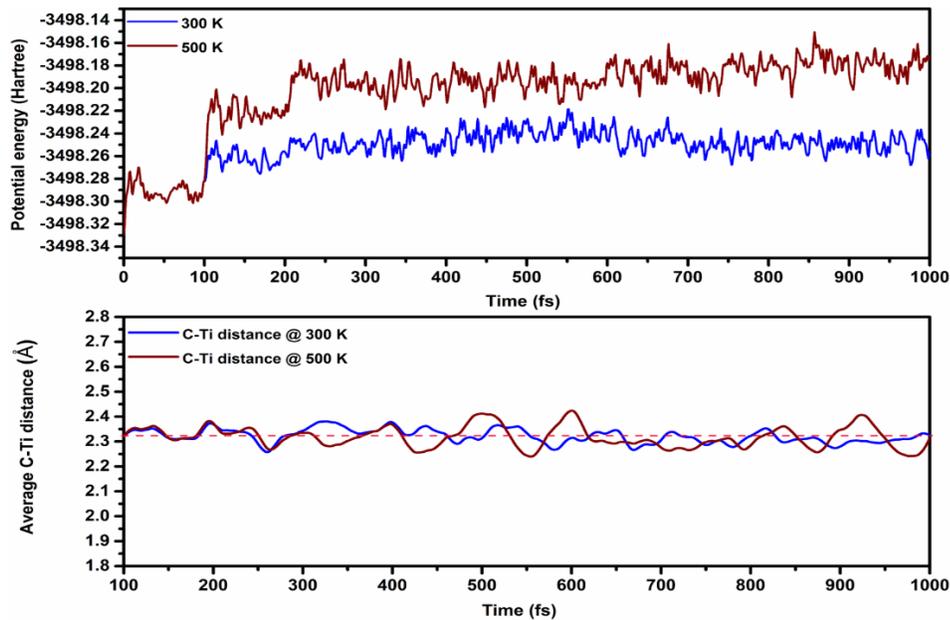

**Figure 11:** (a) Potential energy trajectories of hydrogenated PCP222-3Ti and (b) Time evolution trajectory of average bond length between the Ti atom and C atoms of PCP222 at 300K and 500K.

We have performed molecular dynamic (MD) simulations using the atom-centered density matrix propagation (ADMP) to check the desorption of hydrogen from the PCP222-3Ti-nH$_2$ and the structural integrity of the host. During the simulations, the temperature was maintained by the velocity scaling method, and the temperature was checked and corrected at every time step of 10 fs. Figure 11(a) and S5, show the time variation potential energy trajectories and system snapshots, respectively. The MD simulations at 300K and 1 ps reveal that 2H$_2$ molecules from each Ti atom fly away, and each Ti continues to hold three physisorbed H$_2$ molecules and two chemisorbed hydrogen atoms. When the temperature is elevated to 500 K, almost all the H$_2$ molecules get desorbed and each sorption center hold one physisorbed H$_2$ and two chemisorbed H atoms. Since the first physisorbed H$_2$ is bound strongly with the host material, it may desorb at a higher temperature and time scale. This indicates that the system PCP222-3Ti is not complete reversible at normal temperatures and may show 100% desorption at a higher temperature.

For a practical hydrogen storage material, it is necessary that the host material must keep the structural integrity above the average desorption temperature. To examine the structural integrity of the host material (PCP222-3Ti), we carried out the MD simulations with the host material at 300 K and significantly above the room temperature (500 K) using ADMP. With a time step of 1 fs, the ADMP-MD simulations are carried out for 1 ps. Figure 11(b) depicts the time variation trajectory of the average distance between the Ti atom and the carbon atoms of PCP222 benzene rings. We observe that the PCP222-3Ti maintains its structural stability at 500 K, and the distances between the C-C and C-H bonds essentially remain unchanged. The time evolution trajectories of the average distance between the Ti and C atom of PCP222 were noticed to oscillate about the mean value (2.32 Å) with little variance. This illustrates that the host material's structural stability is maintained significantly above room temperature. In light of this, we believe that PCP222-3Ti can be a viable option for hydrogen storage material.

## 4 Conclusion

In this study, we investigated the thermodynamical stability and hydrogen storage properties of Ti-functionalized [2,2,2]paracyclophane, using the density functional theory. The Ti atoms are strongly bonded to the PCP222 via Dewar mechanism, and no clustering of Ti atoms over PCP222 was noticed. The first H$_2$ molecule is chemisorbed with binding energy of 1.797 eV, while the remaining H$_2$ molecules are physisorbed with an average H$_2$ adsorption energy in the range of 0.467 - 0.953 eV/H$_2$. On saturation with the H$_2$, the Ti atom on PCP222 could adsorb up to 6H$_2$ molecules, while the Ti-2H on PCP222 could adsorb up to 4H$_2$. The average H-H bond distance elongated by 0.06-0.14 Å during the adsorption process which implied that the adsorbed H$_2$ molecules were in quasi-molecular form and the fact is supported by the Hirshfeld charge distribution analysis. . When three Ti atoms were functionalized on PCP222, the H$_2$ gravimetric capacity of the system was up to 7.37 wt%, which was fairly above the US-DOE requirements for practical hydrogen applications. During saturation of H$_2$ adsorption, the host material displayed no significant change in geometry. The thermodynamic usable hydrogen capacity was found to be up to 5 wt% throughout a temperature range of 120-300 K and a pressure range of 3-60 bar. At the temperature of 400 K and pressure of 1-10 bar, the system retains 1.31 wt% of hydrogen which could be desorbed at very high temperatures. A total gravimetric density of up to 6.06 wt% H$_2$ molecules are usable under ambient conditions which is fairly higher than the US-DOE target. MD simulations at 500 K revealed the structural integrity and reversibility of the host and also showed that chemisorbed hydrogens are retained at this temperature. Since, there is no experimental works reported on Ti-functionalized PCP222 for hydrogen storage, we hope our computational work will contribute significantly to the research of hydrogen storage in macrocyclic compounds and provide supporting reference for the future experiments.